\documentclass[12pt]{iopart}
\pdfoutput=1 
  \expandafter\let\csname equation*\endcsname\relax
  \expandafter\let\csname endequation*\endcsname\relax
\usepackage{amsfonts,amstext,amsmath,amssymb,amsthm,graphicx,eucal,float,cite,subfigure}
\newcommand{\ii}{{\rm i}}

\begin{document}
\title{Free Bosons  with a Localized Source}
\author{P. L. Krapivsky$^{1,2}$, Kirone Mallick$^3$ and Dries Sels$^{4,5}$}
\address{$^1$ Department of Physics, Boston University, Boston, Massachusetts 02215 USA}
\address{$^2$ Theoretical Division and CNLS, Los Alamos National Laboratory, 
Los Alamos, New Mexico 87545, USA}
\address{$^3$ Institut de Physique Th\'eorique, Universit\'e Paris-Saclay,
CEA and CNRS, 91191 Gif-sur-Yvette, France}
\address{$^4$ Theory of quantum and complex systems, Universiteit Antwerpen, B-2610 Antwerpen, Belgium}
\address{$^5$ Department of Physics,  Harvard  University, Cambridge, Massachusetts 02138, USA}

\begin{abstract}
We analyze the time evolution of an open quantum system driven by a localized source of bosons. 
We consider non-interacting identical bosons that are injected into a single lattice site and and
perform a continuous 
time quantum walks on a lattice. We show that the average number of bosons grows exponentially 
with time when the input rate  exceeds a certain lattice-dependent critical value. Below the threshold, 
the growth is quadratic in time, which is still much faster than the naive linear in time growth. 
We compute the critical input rate for hyper-cubic lattices and find that it is positive in all
dimensions $d$
except for  $d=2$ where the critical input rate vanishes---the growth is always exponential in two dimensions. To understand the exponential growth, we construct an explicit microscopic Hamiltonian model which gives rise to the open system dynamics once the bath is traced out. Exponential growth is identified with a region of dynamic instability of the Hamiltonian system.
\end{abstract}

\eads{\mailto{pkrapivsky@gmail.com}, ~\mailto{kirone.mallick@ipht.fr} ~ \mailto{dsels@g.harvard.edu}}


\section{Introduction}

We investigate the behavior of an open quantum system of non-interacting identical bosons driven 
by a localized source of bosons. We start with an empty system and assume that
at time $t=0$  the source is turned on and identical bosons are injected at a constant rate at the origin. 

The evolution of open quantum systems is described by the Lindblad equation \cite{Sud76,Lindblad76,Breuer,Manzano}
\begin{equation}
\label{LindbladB}
\partial_t \rho = -\ii [H, \rho] + 2L \rho L^{\dagger}-\{L^{\dagger}L, \rho\}
\end{equation}
for the density matrix $\rho(t)$. The curly brackets in Eq.~\eqref{LindbladB} denote the anti-commutator, 
$\ii = \sqrt{-1}$ and we set $\hbar=1$. The last two terms on the right-hand side of \eqref{LindbladB} represent a shorthand writing of a sum $\sum_{\alpha}\left[2L_\alpha \rho L_\alpha^{\dagger}-\{L_\alpha^{\dagger}L_\alpha, \rho\}\right]$, with operators $L_\alpha$ describing e.g. interactions with reservoirs;  more generally, $\alpha$ labels various dissipative channels. The Lindblad equation defines the general Markovian time evolution, and it has found applications in numerous areas of quantum physics, e.g. in quantum optics \cite{Q:optics,Haroche} and quantum information theory \cite{Nielsen,Preskill}. Extensions of the Lindblad equation description involving more general Markovian quantum stochastic dynamics (see e.g. \cite{Michel,Ohad}) as well as non-Markovian dynamics (see \cite{Non:Markov} for a review) have also been investigated. 

Open quantum systems with time-independent  Lindblad operators have been often studied. In such conditions, a unique stationary limiting state $\rho^\text{ss}=\rho(\infty)$ is reached and the relaxation to this state is well understood \cite{Spohn77}. For periodic driving the system usually approaches a unique limit cycle, see e.g. \cite{Kosloff13,US16}. We consider a quench-type setup in which Lindblad operators suddenly change at a certain moment. The stationary state is never reached in our system, e.g. the total number of particles diverge as $t\to\infty$. To avoid unnecessary complications we analyze the situation when bosons are injected into a single site. Thus we have a single Lindblad operator
\begin{equation}
\label{Lindblad:origin}
L = 
\begin{cases}
0  & t<0\\
\sqrt{\Gamma}\, b_{\bf 0}^{\dagger}  & t>0
\end{cases}
\end{equation}
modeling the input at the origin with intensity $\Gamma$ that is turned on at time $t=0$. 
Equations \eqref{LindbladB}--\eqref{Lindblad:origin} provide a mathematical framework for 
a sudden quench protocol, but instead of the change of a Hamiltonian or temperature which 
is usually investigated the Lindblad operator is changed in our model. For fermions, this sudden 
quenched protocol has been studied in \cite{KMS19} in the lattice realization and in \cite{Spohn10,Kollath} in the continuous setting. For bosons, the problem of localized losses has been investigated both theoretically~\cite{Froml,barmettler,zezyulin,kordas,Sirker,sels} and experimentally~\cite{ott1,ott2}, and it is dual to the problem we consider here as instead of a sink we consider a source. For fermions there is essentially no difference between a source and a sink because a particle sink is the same as a source of holes. However, for bosons a particle source and sink behave in a dramatically different way. While a phase transition can be observed in the case of losses~\cite{ott1,sels}, such a transition is driven by a competition between interaction and dissipation. Here we show there is a rather dramatic phase transition even for free bosons if one considers an incoherent source, rather than a sink. 

We consider free lattice bosons, so the Hamiltonian appearing in Eq.~\eqref{LindbladB} admits a simple expression,
$H =  \sum_{\langle {\bf i}, {\bf j} \rangle}( b_{\bf i}^{\dagger}  b_{\bf j} + b_{\bf j}^{\dagger}  b_{\bf i})$, through
creation and annihilation operators, where the sum is taken over pairs of neighboring lattice sites ${\bf i}$ and ${\bf j}$. 
The non-trivial commutators are  $[ b_{\bf i},  b_{\bf j}^{\dagger}  ] = \delta_{{\bf i},{\bf j}}$,  
all  other  commutators vanish. 
We focus on uniform systems, so the hopping rates are equal and we set them to unity. 

A basic characteristic of this  open quantum system is the total number $\mathcal{N}(t)$
of bosons at time $t$. The full statistics of this evolving random quantity is described by 
the probability distribution $P(N,t)=\text{Prob}[\mathcal{N}(t)=N]$. The exploration
of the full statistics is a challenging problem which is left for the future; here 
we focus on the average total number of bosons
$N(t) = \langle \mathcal{N}(t)\rangle$ and on the density of bosons. 
These characteristics are encoded 
in the two-point correlation function
\begin{equation}
\label{sigmaB:def}
 \sigma_{{\bf i},{\bf j}}(t) = \langle b_{\bf i}^{\dagger}  b_{\bf j}  \rangle = {\rm Tr}\left(\rho(t) b_{\bf i}^{\dagger}  b_{\bf j} \right)
\end{equation}

We limit ourselves to  hyper-cubic lattices $\mathbb{Z}^d$, so we have just two 
dimensionless parameters, $\Gamma$ and $d$. We focus on the long time behavior, $t\gg 1$. 
One anticipates a linear growth of $N(t)$, more precisely 
\begin{equation}
\label{N:naive}
N_\text{naive}(t) = 2\Gamma t
\end{equation}
This  naive growth law would be exact in arbitrary dimension if non-interacting particles were 
distinguishable. Quantum statistics induces effective repulsion of identical fermions 
and attraction of identical bosons, so one expects \eqref{N:naive} to provide an upper bound for fermions and a lower bound for bosons. 
This is indeed correct. Furthermore, we have recently shown \cite{KMS19} that the asymptotic growth 
is linear in time for fermions. More precisely, $N(t) = c_d(\Gamma) t$ with $c_d(\Gamma)<2\Gamma$
and  $\Gamma^{-1}c_d(\Gamma)\to 2$ as $\Gamma\to+0$. For bosons, in contrast, 
the lower bound \eqref{N:naive} is essentially useless as the actual asymptotic growth of $N(t)$ 
is super-linear. Below we show that this growth is remarkably rich: A dynamical phase transition 
between algebraic and exponential growth occurs when the input rate passes through the critical value. 
In one dimension, the critical input rate is $\Gamma_1=2$ and 
\begin{equation}
\label{N1d:all}
N(t)\simeq
\begin{cases}
A_1(\Gamma)\,e^{2t\sqrt{\Gamma^2-4}}  & \Gamma>2 \\
\frac{64}{3}\,t^4                                         & \Gamma = 2  \\
B_1(\Gamma)\, t^2                                   & \Gamma < 2  \\
\end{cases}
\end{equation}
with amplitudes
\begin{subequations}
\begin{align}
\label{A1}
A_1(\Gamma) & = \frac{\Gamma^4}{(\Gamma^2-4)^2}\\
\label{B1}
B_1(\Gamma) & = \frac{4\Gamma^4}{4-\Gamma^2}
\end{align}
\end{subequations}

When $d\geq 3$, the asymptotic behaviors of the average total number of bosons are rather similar to
that for the one-dimensional case:
\begin{equation}
\label{Nd:all}
N(t)\simeq
\begin{cases}
A_d(\Gamma)\,e^{2C_d(\Gamma)\, t}  & \Gamma>\Gamma_d \\
B_d^\text{crit}\,t^2                                 & \Gamma = \Gamma_d  \\
B_d(\Gamma)\, t^2                               & \Gamma < \Gamma_d  \\
\end{cases}
\end{equation}
The only difference with the one-dimensional growth laws represented by Eq.~\eqref{N1d:all} 
is that the time dependence in the critical regime is the same as in the sub-critical regime, $N\propto t^2$; 
the amplitude undergoes a phase transition, $B_d^\text{crit}\ne B_d(\Gamma_d)$.

The two-dimensional case is exceptional: The critical input rate vanishes, $\Gamma_2=0$, 
and hence the exponential growth, $N\simeq A_2(\Gamma)\,e^{2C_2(\Gamma)\, t}$, 
occurs for all $\Gamma>0$. The critical input rate seems to  vanish for all two-dimensional 
lattices,  reflecting the peculiar role of $d=2$. 

In section \ref{sec:1d} we analyze the behavior in one dimension and establish the growth law \eqref{N1d:all}. 
The density profile in one dimension is derived in section \ref{sec:1d-density}. Section \ref{sec:high-B} analyzes the behavior in higher dimensions. The density profile in higher dimensions is investigated in section \ref{sec:d-density}. An explicit microscopic Hamiltonian model for the system, including a bath, is constructed in section~\ref{sec:model}. Conclusions are given in section \ref{sec:concl}.  

\section{One dimension: Total number of bosons}
\label{sec:1d}

In one dimension, the Hamiltonian reads 
\begin{equation}
\label{HamB}
H =  \sum_{n=-\infty}^\infty \left( b_n^{\dagger}  b_{n+1} +
b_{n+1}^{\dagger} b_{n} \right)
\end{equation} 

An exact solution for the density matrix $\rho$ is an intriguing challenge which is left to future work. Here we focus on the two-point correlation function 
\begin{equation}
\label{sigma:def}
 \sigma_{i,j}(t) =  {\rm Tr}\left(\rho(t) b_i^{\dagger} b_j \right)
\end{equation}
Using the Lindblad equation \eqref{LindbladB} with Hamiltonian \eqref{HamB} 
and $L=\sqrt{\Gamma}\, b_0^{\dagger}$ we deduce a closed system of coupled differential equations
\begin{eqnarray}
\label{sigmaB}
 \frac{d  \sigma_{i,j}}{dt} &=& \ii \left( \sigma_{i+1,j} +
 \sigma_{i-1,j} - \sigma_{i,j+1} -  \sigma_{i,j-1} \right) + \Gamma\left(\delta_{i,0} \sigma_{i,j}+ \delta_{j,0}
 \sigma_{i,j}\right) \nonumber \\
 &+&  2\Gamma\delta_{i,0} \delta_{j,0}
\end{eqnarray}
for the two-point correlation function. This infinite system of linear equations admits an exact solution. To derive this solution we first drop the source term, viz. the term in the second line in \eqref{sigmaB}, solve the resulting {\em homogeneous} system of equations, and then express the solution of \eqref{sigmaB} through the solution of the homogeneous system. Thus we start with 
\begin{equation}
\label{sigma-no}
 \frac{d  \sigma_{i,j}}{dt} = \ii \left( \sigma_{i+1,j} +
 \sigma_{i-1,j} - \sigma_{i,j+1} -  \sigma_{i,j-1} \right)
 + \Gamma\left(\delta_{i,0} \sigma_{i,j}+ \delta_{j,0}
 \sigma_{i,j}\right)
\end{equation}
The factorization ansatz
\begin{equation}
  \sigma_{i,j}(t) = S_i(t) S_j^*(t)
  \label{2pt-ansatz}
\end{equation}
is consistent with \eqref{sigma-no} if the functions $S_n(t)$ satisfy 
\begin{equation}
\label{S-dual}
 \frac{d  S_n}{dt} = \ii [S_{n+1} +  S_{n-1}] +  \Gamma \delta_{n,0} S_n
\end{equation}
Similar equations have been analyzed in \cite{KLM14,KMS19}; below we employ the same procedure.  

A proper choice of the initial condition which will allow us to treat the original system \eqref{sigmaB} starting from the empty lattice is 
\begin{equation}
\label{IC}
S_n(0) = \delta_{n,0}
\end{equation}
The solution of the initial-value  problem \eqref{S-dual}--\eqref{IC} can be obtained 
using the Laplace-Fourier transform. The Laplace transform with respect to time,
\begin{equation}
\widehat{S}_n(s) = \int_0^\infty dt\,e^{-st} S_n(t) \, , 
\end{equation}
recasts \eqref{S-dual} into
\begin{equation}
\label{S-Lap}
s\widehat{S}_n - \delta_{n,0} = \ii \big[\widehat{S}_{n+1} +
  \widehat{S}_{n-1}\big] + \Gamma \delta_{n,0} \widehat{S}_n
\end{equation}
Performing the Fourier transform in space, 
\begin{equation}
\label{S-Fourier}
S(s,q) =  \sum_{n=-\infty}^\infty \widehat{S}_n(s)\,e^{- \ii qn}\,, 
\end{equation}
we obtain
\begin{equation}
\label{SS0}
S(s,q) = \frac{1 +  \Gamma \widehat{S}_0(s)}{s - 2 \ii \cos q}
\end{equation}
The definition \eqref{S-Fourier} implies 
\begin{equation*}
\widehat{S}_0(s) = \frac{1}{2\pi} \int_0^{2\pi} dq\,S(s,q)
\end{equation*}
which in conjunction with  \eqref{SS0} fixes 
\begin{equation}
\label{S0}
\widehat{S}_0(s) = \frac{1}{\sqrt{s^2+4} -\Gamma}
\end{equation}
and yields 
\begin{equation}
\label{S-sol}
S(s,q) = \frac{\sqrt{s^2+4}}{\sqrt{s^2+4} -\Gamma} \,\, \frac{ 1 } {s - 2 \ii \cos q}
\end{equation}
Inverting the Fourier transform we arrive at 
\begin{equation}
\label{Sn-Lap}
\widehat{S}_n(s) = \frac{1}{\sqrt{s^2+4} -\Gamma}\,
\left(\frac{2\ii}{s+\sqrt{s^2+4}}\right)^{|n|}
\end{equation}

Returning to the original problem we note that Eqs.~\eqref{sigmaB} can be rewritten as
\begin{equation}
  \dot \sigma = M  \sigma + V
\label{sigma:MV}
\end{equation}
where $\sigma=\sigma(t)$ is an infinite vector with components $\sigma_{i,j}(t)$, 
the matrix $M$ encodes the homogeneous terms, and the source term $V$ is a vector 
with a unique non-zero component,  $2 \Gamma \delta_{i,0} \delta_{j,0}$. Solving \eqref{sigma:MV} gives
\begin{equation}
   \sigma(t) = \int_0^t {\rm d}\tau \, e^{(t - \tau) M} V
\end{equation}
Noting that $e^{t M}$ is the solution of the initial-value  problem \eqref{S-dual}--\eqref{IC}
and using the ansatz \eqref{2pt-ansatz} we establish 
\begin{equation}
\label{sigma:S}
  \sigma_{i,j}(t)  =   2 \Gamma \int_0^t {\rm d}\tau  S_i(\tau) S_j^*(\tau)
\end{equation}
with functions $S_i$ satisfying  the dynamics \eqref{S-dual}. 

The average total number of particles is 
\begin{equation}
\label{N:sigma}
N(t) =  \sum_{n=-\infty}^\infty \sigma_{n,n}(t) 
\end{equation}
Combining this result with \eqref{sigma:S} we obtain 
\begin{equation}
\label{N1}
 N(t) = 2 \Gamma \int_0^t {\rm d}\tau \, \sum_{n=-\infty}^\infty | S_n(\tau)|^2
\end{equation}
Using \eqref{S-dual} we express the sum in \eqref{N1} through $S_0(t)$:
\begin{equation}
\label{sum}
\sum_{n=-\infty}^\infty |S_n(t)|^2= 1 + 2\Gamma \int_0^t dt'\, |S_0(t')|^2
\end{equation}
Differentiating \eqref{N1} and using \eqref{sum} we arrive at
\begin{equation}
\label{N1:S0}
\frac{d N}{dt} = 2\Gamma\left[1 + 2\Gamma \int_0^t dt'\, |S_0(t')|^2\right]
\end{equation}
from which $N(t)\geq 2\Gamma t$. Thus \eqref{N:naive} indeed provides the lower bound as we have asserted. 
Differentiating \eqref{N1:S0} we obtain
\begin{equation}
\label{N:S0}
\frac{d^2 N}{dt^2} = (2\Gamma)^2\, |S_0(t)|^2
\end{equation}
which we shall also use below. 

The Laplace transform \eqref{S0} has a neat form, but inverting it in terms of special functions 
appears impossible. A rather simple integral representation of $S_0(t)$ is possible, however. 
To derive it we rely on a useful general identity for inverse Laplace
transforms. Suppose we know the inverse Laplace transform $f(t)$  of
$\widehat{f}(s)$, but we actually want to determine the inverse Laplace
transform $F(t)$ of $\widehat{F}(s) \equiv \widehat{f}\big(\sqrt{s^2+a^2}\big)$. This inverse Laplace
transform admits \cite{Bateman:Lap} a general 
expression through $f(t)$,
\begin{equation}
\label{inv-Lap-root}
F(t) = f(t) - a\int_0^t d\tau\, J_1(a\tau)\,f\big(\sqrt{t^2-\tau^2}\big),
\end{equation}
where $J_1$ is the Bessel function. Specializing to \eqref{S0} we obtain
\begin{equation}
\label{S0:int-rep}
S_0(t)=e^{\Gamma t} - 2\int_0^t d\tau\, J_1(2\tau)\,e^{\Gamma\sqrt{t^2-\tau^2}}
\end{equation}

The integral representation \eqref{S0:int-rep} is useful for extracting the small time behavior: 
\begin{equation*}
S_0(t)=1 + \Gamma t + \tfrac{1}{2}(\Gamma^2-2)t^2+ \tfrac{1}{6}\Gamma(\Gamma^2-4)t^3
+ \tfrac{1}{24}(\Gamma^4 - 6\Gamma^2 + 6)t^4+\ldots
\end{equation*}
Combining this expansion with \eqref{N1:S0} we obtain
\begin{eqnarray}
N &=& 2\Gamma t + 2\Gamma^2 t^2 + \tfrac{4}{3} \Gamma^3 t^3 + \tfrac{2}{3}\Gamma^2(\Gamma^2-1)t^4 \nonumber\\
&+& \tfrac{2}{15}\Gamma^2(2\Gamma^2-5)t^5 + \tfrac{1}{45}\Gamma^2(4\Gamma^4-17\Gamma^2+9)t^6+\ldots
\end{eqnarray}
The leading small time asymptotic is linear in time in agreement with intuition. 
The lower bound \eqref{N:naive} actually provides the leading small time asymptotic. 

The large time growth is faster than linear as we show below. 
The leading behavior in the $t\to\infty$ limit is easier to extract directly from 
the Laplace transform \eqref{S0} than from the integral representation \eqref{S0:int-rep}. 

\subsection{The supercritical regime:  $\Gamma > 2$}
 
To determine the leading large time behavior of $S_0(t)$ when the source is sufficiently strong, $\Gamma > 2$, it is useful to re-write \eqref{S0} as 
\begin{equation}
\label{S0:2}
\widehat{S}_0(s) = \frac{1}{s - \sqrt{\Gamma^2-4}} \,\frac{\sqrt{s^2+4} +\Gamma}{s+\sqrt{\Gamma^2-4}}
\end{equation}
This Laplace transform has a simple pole at $s=\sqrt{\Gamma^2-4}$ with residue  
$\Gamma/\sqrt{\Gamma^2-4}$. Therefore the asymptotic growth of $S_0(t)$ is exponential  
\begin{equation}
\label{S0:1d+}
S_0(t)\simeq \frac{\Gamma}{\sqrt{\Gamma^2-4}}\,e^{t\sqrt{\Gamma^2-4}}
\end{equation}
and the total number of bosons found from \eqref{N:S0} and \eqref{S0:1d+} is also exponential 
\begin{equation}
\label{N1d:super}
N(t)\simeq \frac{\Gamma^4}{(\Gamma^2-4)^2}\, e^{2t\sqrt{\Gamma^2-4}}
\end{equation}
This is the announced asymptotic, \eqref{N1d:all}  and \eqref{A1}, in the $\Gamma>2$ region.

\begin{figure}[t]
	\centering
\subfigure[]{\includegraphics[width=0.45\textwidth]{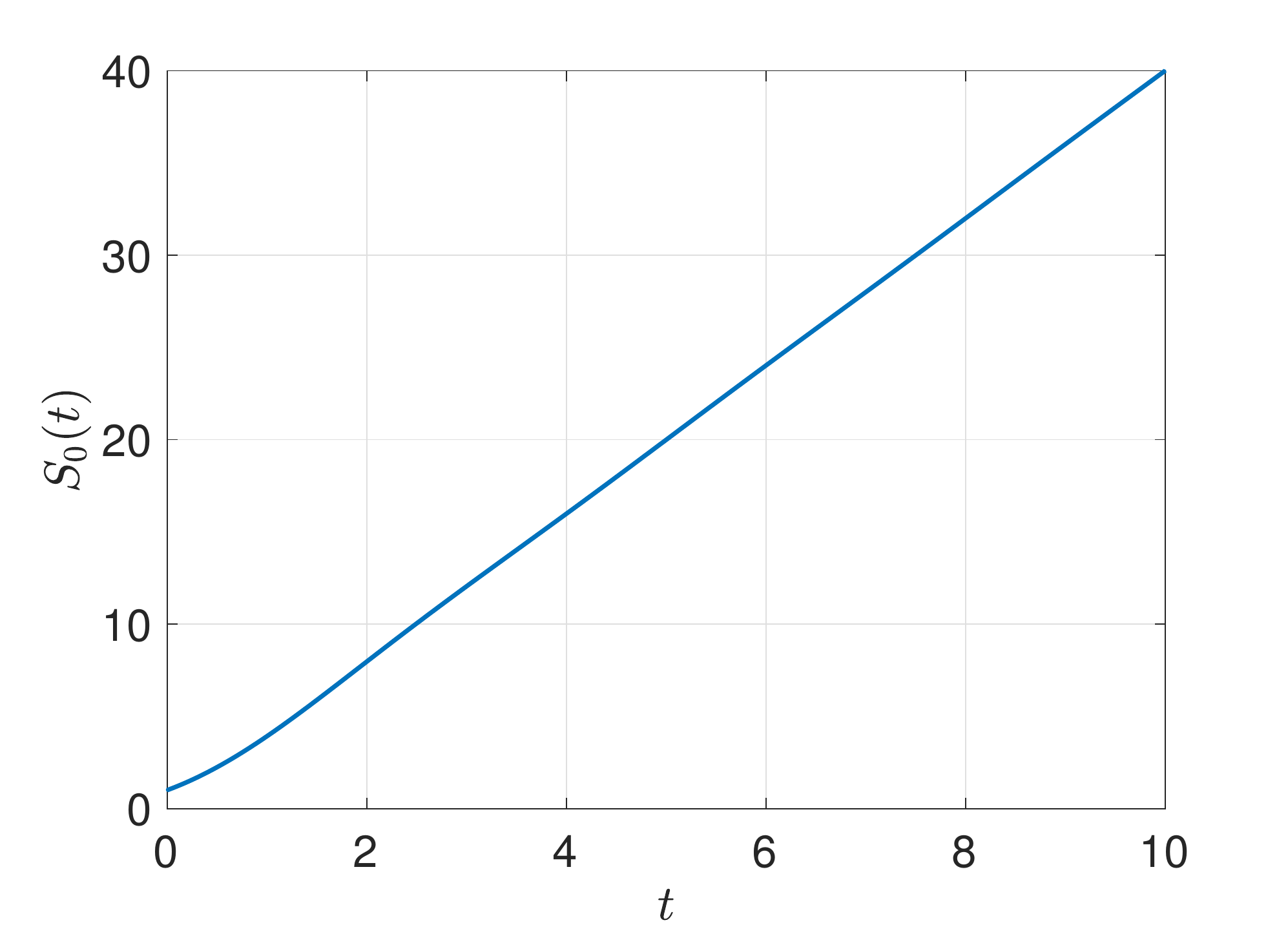}}\qquad
\subfigure[]{\includegraphics[width=0.45\textwidth]{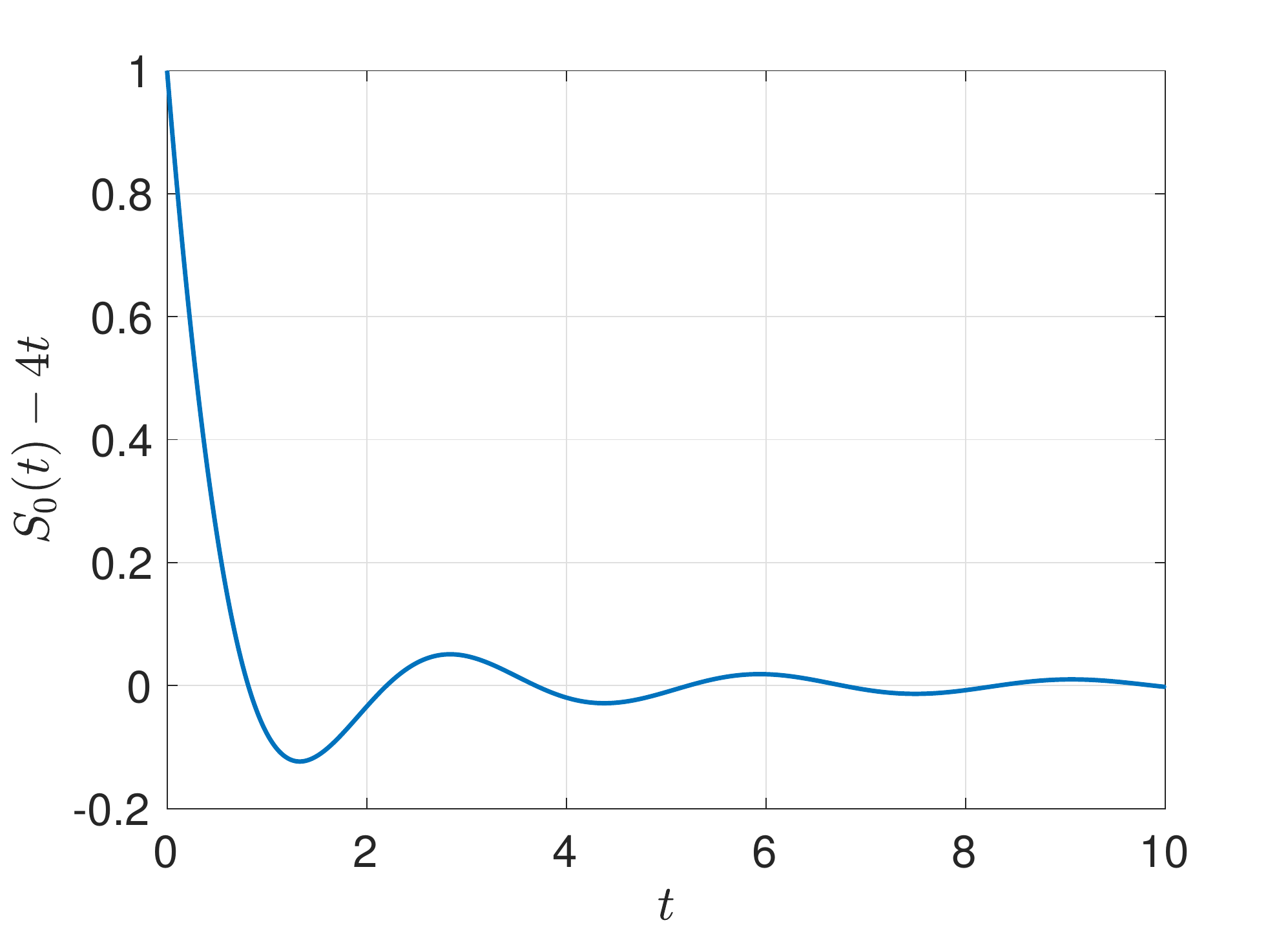}}
	\caption{(a) $S_0(t)$ versus time $t$; we compute $S_0$ using the integral representation \eqref{S0:int-rep} with $\Gamma=2$. 
	(b) The deviation $S_0(t) - 4t$ from the leading asymptotic approaches to zero 
	in an oscillatory manner. }
	\label{fig:S0}
\end{figure}

\subsection{The critical case:  $\Gamma = 2$}

The Laplace transform \eqref{S0:2} simplifies to
\begin{equation}
\label{S0:crit}
\widehat{S}_0(s) = \frac{\sqrt{s^2+4} +2}{s^2}
\end{equation}
and behaves as $4/s^2$ when $s\to 0$, implying that $S_0(t)\simeq 4t$ as $t\to\infty$, see also Fig.~\ref{fig:S0}~(a).
Plugging $S_0(t)\simeq 4t$ into \eqref{N:S0} we obtain the announced asymptotic \eqref{N1d:all} at $\Gamma=2$, namely 
\begin{equation}
\label{N1d:crit}
N(t)\simeq \frac{64}{3}\,t^4 
\end{equation}

Figure \ref{fig:S0}~(b) shows that $S_0(t) - 4t$ approaches to zero in an oscillatory manner. 
To derive this sub-leading correction analytically, we re-write \eqref{S0:crit} as 
\begin{equation}
\label{S0:crit-Lap}
\widehat{S}_0(s) = \frac{4}{s^2}+ \widehat{F}_*(s), \quad  \widehat{F}_*(s)= \frac{1}{\sqrt{s^2+4} +2}
\end{equation}
Performing the inverse Laplace transform and using the identity \eqref{inv-Lap-root} we establish 
an integral representation 
\begin{equation}
\label{S0:crit-time}
S_0(t) - 4t=F_*(t), \quad F_*(t) = e^{-2t} - 2\int_0^t d\tau\, J_1(2\tau)\,e^{-2\sqrt{t^2-\tau^2}}
\end{equation}
The leading long-time behavior is now readily computed:
\begin{equation}
S_0(t) - 4t \simeq -\frac{J_1(2t)}{2t} \simeq \frac{\cos\!\left(2t+\frac{\pi}{4}\right)}{\sqrt{4\pi t^3}} 
\end{equation}
for $t\gg 1$. Using \eqref{N1:S0} we find that the sub-leading correction 
to the asymptotic growth law \eqref{N1d:crit} scales linearly with time:
\begin{equation}
\label{N1d:crit-more}
N(t) -  \frac{64}{3}\,t^4 \simeq 4(1+f_*) t, \quad f_*=4\int_0^\infty dt\,F_*(t)[F_*(t)+8t]
\end{equation}

\subsection{The subcritical regime:  $0 < \Gamma < 2$}

To determine the asymptotic behavior of $S_0(t)$ when the source is sufficiently weak, $\Gamma < 2$, 
it is useful to re-write \eqref{S0} as 
\begin{equation}
\label{S0:1}
\widehat{S}_0(s) = \frac{\sqrt{s^2+4} +\Gamma}{s^2+4-\Gamma^2}
\end{equation}
This Laplace transform has poles at $s=\pm \ii \sqrt{4-\Gamma^2}$. 
The contribution of these poles to the inverse Laplace transform 
\begin{equation}
\label{S0:1-osc}
S_0(t)\simeq \frac{2\Gamma}{\sqrt{4-\Gamma^2}}\,\sin\!\left(t \sqrt{4-\Gamma^2}\right)
\end{equation}
is asymptotically dominant. Therefore
\begin{equation}
\label{int:S0}
\int_0^t dt'\, |S_0(t')|^2 \simeq \frac{2\Gamma^2}{4-\Gamma^2}\,\,t
\end{equation}
which in conjunction with \eqref{N1:S0} leads to the announced asymptotic, 
\eqref{N1d:all}  and \eqref{B1}, in the $\Gamma<2$ region. 

The sub-leading correction is determined using the same procedure as in the critical case, cf. \eqref{S0:crit-Lap}--\eqref{S0:crit-time}. Namely we write
\begin{equation}
\label{S0:subcrit-Lap}
\widehat{S}_0(s) = \frac{2\Gamma}{s^2+4-\Gamma^2}+ \widehat{F}(s), \quad  
\widehat{F}(s)= \frac{\sqrt{s^2+4} -\Gamma}{s^2+4-\Gamma^2}
\end{equation}
from which we deduce an integral representation of the sub-leading correction 
\begin{equation}
\label{F:subcrit}
F(t) = e^{-\Gamma t} - 2\int_0^t d\tau\, J_1(2\tau)\,e^{-\Gamma\sqrt{t^2-\tau^2}}
\end{equation}
which leads to
\begin{equation}
\label{S0:subcrit-time}
S_0(t) -  \frac{2\Gamma}{\sqrt{4-\Gamma^2}}\,\sin\!\left(t \sqrt{4-\Gamma^2}\right)
\simeq \frac{2}{\Gamma^2}\,\frac{\cos\!\left(2t+\frac{\pi}{4}\right)}{\sqrt{\pi t^3}} 
\end{equation}
Straightforward calculations lead to the following more accurate asymptotic expansion
for the average number of bosons,
\begin{equation}
\label{N1d:subcrit-more}
N(t) = B_1(\Gamma)\,t^2 + 2\Gamma [1+f(\Gamma)]t+ \frac{4\Gamma^4}{(4-\Gamma^2)^2}\,\cos\!\left(2t \sqrt{4-\Gamma^2}\right),
\end{equation}
with
\begin{equation}
\label{f-Gamma}
f(\Gamma)=2\Gamma\int_0^\infty dt\,F^2(t) + \frac{8\Gamma^2}{\sqrt{4-\Gamma^2}}
\int_0^\infty dt\,F(t) \sin\!\left(t \sqrt{4-\Gamma^2}\right)
\end{equation}
where $F(t)$ is given by \eqref{F:subcrit}.

\section{Density profile in one dimension}
\label{sec:1d-density}

Let us first determine the density at the origin $\rho_0(t)\equiv \sigma_{0,0}(t)$. Equation \eqref{sigma:S} shows that
\begin{equation}
\label{n0}
\rho_0(t) = 2\Gamma \int_0^t dt'\, |S_0(t')|^2
\end{equation}
Using already established results for $S_0(t)$ we arrive at the asymptotic behaviors
\begin{equation}
\label{N0-1}
\rho_0(t)\simeq
\begin{cases}
a_1(\Gamma)\,e^{2t\sqrt{\Gamma^2-4}}  & \Gamma>2 \\
\frac{64}{3}\,t^3                                         & \Gamma = 2  \\
b_1(\Gamma)\, t                                       & \Gamma < 2  \\
\end{cases}
\end{equation}
with amplitudes
\begin{subequations}
\begin{align}
\label{a1}
a_1(\Gamma) & = \left(\frac{\Gamma}{\sqrt{\Gamma^2-4}}\right)^3\\
\label{b1}
b_1(\Gamma) & = \frac{4\Gamma^3}{4-\Gamma^2}
\end{align}
\end{subequations}

A more accurate long time behavior of the density at the origin can be established using 
results of the previous section. When $\Gamma=2$
\begin{equation}
\rho_0(t)=\frac{64}{3}\,t^3  + f_* + O(t^{-1/2})
\end{equation}
with $f_*$ appearing in \eqref{N1d:crit-more}, while for $\Gamma<2$
\begin{equation}
\rho_0(t)=b_1(\Gamma)\, t   
- 2\left(\frac{\Gamma}{\sqrt{4-\Gamma^2}}\right)^3 \sin\!\left(2t \sqrt{4-\Gamma^2}\right) + f(\Gamma) + O(t^{-3/2})
\end{equation}
with $f(\Gamma)$ given by \eqref{f-Gamma}. 

We now analyze the entire density profile.

\subsection{The supercritical case: $\Gamma > 2$}
 
To determine the asymptotic behavior of $S_n(t)$ in the super-critical regime, $\Gamma > 2$, we guess that $S_n(t)$ exhibits the  same exponential growth as $S_0(t)$, see \eqref{S0:1d+}, that is
\begin{equation}
\label{Sn:1d+}
S_n(t)\simeq F_n\,e^{t\sqrt{\Gamma^2-4}}
\end{equation}
Plugging this ansatz into \eqref{S-dual} we obtain
\begin{equation}
\label{Fn:rec}
\sqrt{\Gamma^2-4}\,F_n = \ii \left[F_{n+1}+F_{n-1}\right]
\end{equation}
for $n\ne 0$. Solving this simple recurrence yields\footnote{We keep only the solution
  that vanishes  when $|n|\to\infty$.}
\begin{equation}
\label{Fn:sol}
F_n = C\, \ii^{|n|}\left(\frac{\Gamma - \sqrt{\Gamma^2-4}}{2}\right)^{|n|}
\end{equation}
At the origin the recurrence \eqref{Fn:rec} should be modified to 
\begin{equation}
\label{F0:rec}
\sqrt{\Gamma^2-4}\,F_0 = \ii \left[F_{1}+F_{-1}\right]+\Gamma F_0
\end{equation}
One can verify that \eqref{F0:rec} is consistent with the exact solution \eqref{Fn:sol}. 

\begin{figure}[t]
\centering
\includegraphics[width=0.6 \textwidth]{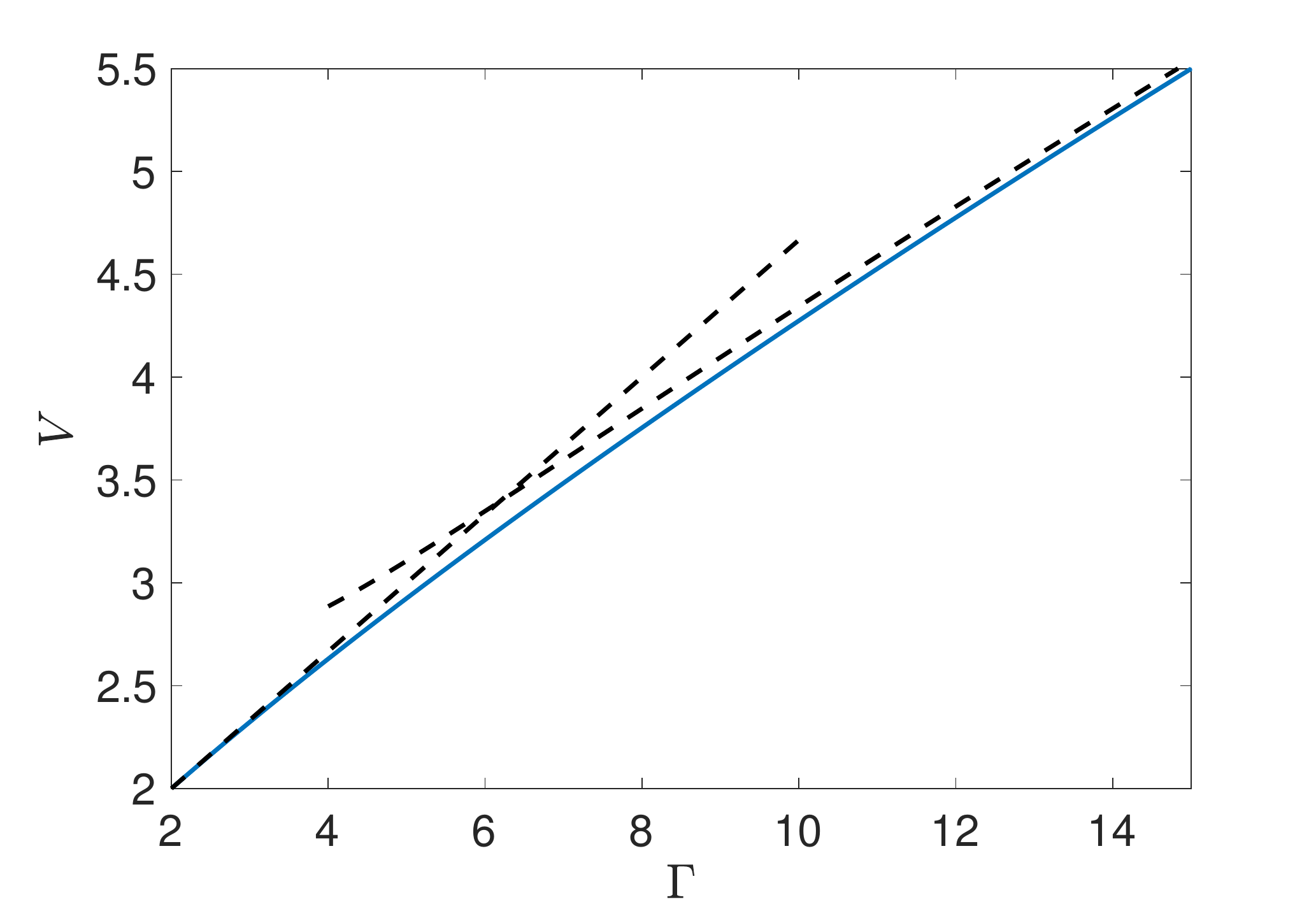}
\caption{The speed $V$ of the extension of the droplet of bosons versus the input rate $\Gamma$ in one dimension in the supercritical regime. The analytical prediction for the speed is given by \eqref{speed}; dashed lines show asymptotic results. }
\label{fig:V-super}
\end{figure}

Recalling \eqref{sigma:S} we have
\begin{equation}
\label{rho:def}
\rho_n(t) \equiv \sigma_{n,n}(t) = 2\Gamma\int_0^t d\tau\, S_n(\tau) S_n^*(\tau)
\end{equation}
Combining this result with \eqref{Sn:1d+} and \eqref{Fn:sol} we compute 
 the density. Matching with already known density at the origin we obtain
\begin{equation}
\label{rho-n:1d+}
\frac{\rho_n(t)}{\rho_0(t)} = \left(\frac{\Gamma - \sqrt{\Gamma^2-4}}{2}\right)^{2|n|}
\end{equation}
Thus the density profile normalized to its value at the origin
is purely exponential in the super-critical regime. Re-writing \eqref{rho-n:1d+} as 
\begin{equation}
\label{rho-n:+}
\rho_n(t) = a_1(\Gamma) \exp\!\left\{2 \ln\frac{\Gamma + \sqrt{\Gamma^2-4}}{2}\,[Vt-|n|]\right\}
\end{equation}
with 
\begin{equation}
\label{speed}
V = \frac{\sqrt{\Gamma^2-4}}{\ln\frac{\Gamma + \sqrt{\Gamma^2-4}}{2}}
\end{equation}
we conclude that the interval $[-Vt,Vt]$ is filled with bosons, and at most few bosons are outside of this interval. The speed $V$ of the extension of the droplet of bosons exceeds the genuine speed, 2 in our units, with which a single continuous time quantum walker would have spread (see Fig.~\ref{fig:V-super}). This excess is caused by the exponential growth of the total number of bosons---the exponentially small spills over the natural $\pm 2t$ boundaries for each quantum walker accumulate to increase the speed to $V(\Gamma)>2$. 

The extremal behaviors of the speed are
\begin{subequations}
\begin{align}
\label{speed-excess-small}
V & =2 + \tfrac{1}{3}(\Gamma-2) - \tfrac{1}{90}(\Gamma-2)^2 +\ldots  \quad (\Gamma\to 2 + 0)\\
\label{speed-excess-large}
V & = \frac{\Gamma}{\ln \Gamma} - \frac{2}{\Gamma\, \ln \Gamma} + \frac{1}{\Gamma (\ln \Gamma)^2}+\ldots  \quad ~(\Gamma\to\infty)
\end{align}
\end{subequations}

\subsection{The subcritical domain: $0 < \Gamma < 2$}

\begin{figure}[t]
\centering
\includegraphics[width=0.77 \textwidth]{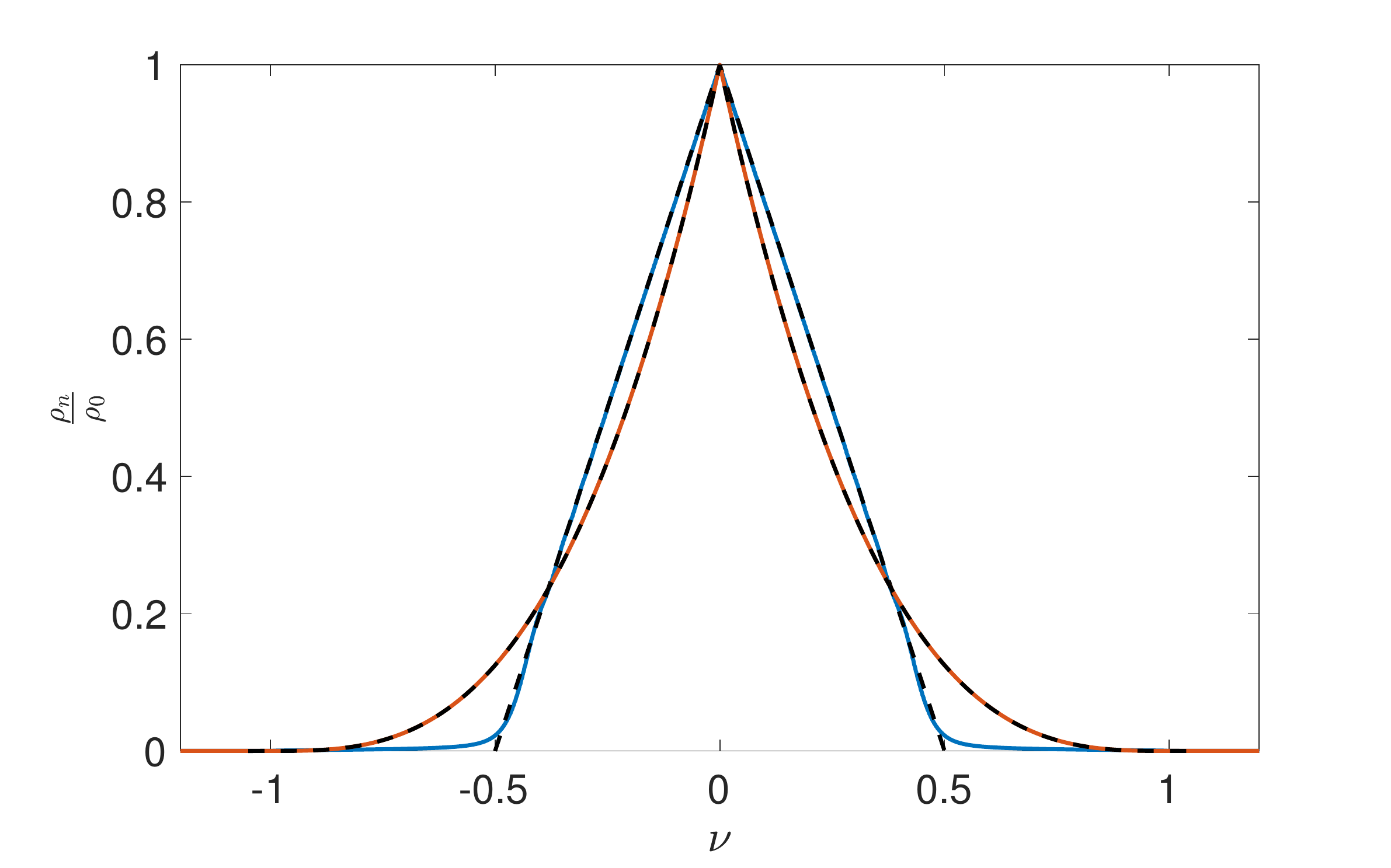}
\caption{The density profile normalized to its value at the origin in the sub-critical regime (the profile at $\Gamma=1$  is shown in blue) and in the critical regime ($\Gamma=2$) in red. The dashed lines show the analytic asymptotic results and the full lines are numeric results computed at $t=400$.}
\label{fig:density-tent}
\end{figure}

In the sub-critical regime, $0 < \Gamma < 2$, we have $\rho_n\sim \rho_0\sim t$ which is consistent with \eqref{rho:def} when 
$S_n=O(1)$. More precisely, $S_n(t)$ is expected to admit a scaling form
\begin{equation}
\label{Sn:1d-}
S_n(t)\simeq \ii^{|n|}\, \mathcal{S}(\nu), \quad \nu=\frac{n}{2t}
\end{equation}
in the scaling limit
\begin{equation}
\label{scaling}
|n|\to\infty, \quad t\to\infty, \quad \nu=\frac{n}{2t}=\text{finite}
\end{equation}
Inserting \eqref{Sn:1d-} into  \eqref{S-dual} and replacing the difference $\mathcal{S}[\nu-1/(2t)]-\mathcal{S}[\nu+1/(2t)]$ by derivative $t^{-1}\frac{d\mathcal{S}}{d\nu}$ we arrive at a simple equation $(1-\nu)\frac{d\mathcal{S}}{d\nu}= 0$ for the scaling function $\mathcal{S}(\nu)$. Therefore $\mathcal{S}(\nu)$ is a constant, more precisely 
\begin{equation}
\label{Phi:-}
\mathcal{S} = C \times 
\begin{cases}
1 & |\nu|<\nu_*\\
0 & |\nu|>\nu_*
\end{cases}
\end{equation}
We expect $\nu_*\leq 1$ since bosons propagate at most on distance $\pm 2t$ from the origin. Plugging \eqref{Phi:-} into \eqref{rho:def} we obtain 
\begin{equation}
\frac{\rho_n}{\rho_0} = 
\begin{cases}
1- |\nu|/\nu_* & |\nu|<\nu_*\\
0                    & |\nu|>\nu_*
\end{cases}
\end{equation}
Using this profile we compute the total number of bosons $N=\rho_0 \nu_* 2t$. Combining this with known asymptotic behaviors 
\begin{equation*}
N =  \frac{4\Gamma^4}{4-\Gamma^2}\, t^2, \quad \rho_0= \frac{4\Gamma^3}{4-\Gamma^2}\, t
\end{equation*}
we fix $\nu_*=\Gamma/2$. Thus in the subcritical regime the asymptotic density profile normalized to its value at the origin is 
\begin{equation}
\label{tent}
\frac{\rho_n}{\rho_0} = 
\begin{cases}
1- 2|\nu|/\Gamma & |\nu|<\Gamma/2\\
0                           & |\nu|>\Gamma/2
\end{cases}
\end{equation}
This  tent-like density profile is shown in Fig.~\ref{fig:density-tent}. 

\subsection{The critical regime: $\Gamma = 2$.}

In the critical regime, $\rho_n\sim \rho_0\sim t^3$ which is consistent with \eqref{rho:def} when $S_n=O(t)$. More precisely, the density admits the scaling form
\begin{equation}
\label{Sn:1d-crit}
S_n(t)\simeq \ii^n\, t\,\mathcal{S}(\nu), \quad \nu=\frac{n}{2t}
\end{equation}
in the scaling limit \eqref{scaling}. Plugging \eqref{Sn:1d-crit} into  \eqref{S-dual} we arrive at the differential equation 
$(1-\nu)\mathcal{S}' = -\mathcal{S}$ which is solved to give\footnote{Our analysis is performed in the region $0<\nu\leq 1$. The solution for $-1\leq \nu<0$ is determined by symmetry. There are essentially no bosons in the $|\nu|>1$ region.}
\begin{equation}
\label{Phi:crit}
\mathcal{S} = C \times 
\begin{cases}
1 - |\nu| & |\nu|< 1\\
0 & |\nu|>1
\end{cases}
\end{equation}
Using \eqref{Sn:1d-crit} and \eqref{Phi:crit} we compute the integral in \eqref{rho:def}
\begin{equation*}
\rho_n  =  4 C^2\int_{n/2}^t d\tau\,\tau^2(1-\nu)^2 = \frac{4 C^2}{3} \left(t-\frac{n}{2}\right)^3
\end{equation*}
Thus the asymptotic density profile normalized to its value at the origin is 
\begin{equation}
\label{cube}
\frac{\rho_n(t)}{\rho_0(t)} = (1-|\nu|)^3, \quad \rho_0(t) = \frac{64}{3}\,t^3 
\end{equation}
in the critical regime. This cubic density profile is shown in Fig.~\ref{fig:density-tent}.

\section{Free bosons on ${\mathbb Z}^d$ with a source}
\label{sec:high-B}

Here we consider non-interacting identical bosons performing independent continuous time quantum walks on ${\mathbb Z}^d$ and coupled to a source at the origin. As in one dimension, we first drop the source term in equations for $\sigma_{{\bf i},{\bf j}}(t)$ and use the factorization ansatz. We then perform the Laplace transform for quantities $S_{\bf n}(t)$
\begin{subequations}
\begin{equation}
\label{Sn-Lap-d}
\widehat{S}_{\bf n}(s) = \int_0^\infty ds\,e^{-st} S_{\bf n}(t)
\end{equation}
followed by the Fourier transform
\begin{equation}
\label{Ssq-d}
S(s, {\bf q}) = \sum_{{\bf n}}  \widehat{S}_{\bf n}(s)\,e^{- \ii {\bf
    q}\cdot {\bf n}}
\end{equation}
\end{subequations}
Here ${\bf n} = (n_1,\ldots,n_d)$, ${\bf q} = (q_1,\ldots,q_d)$
and ${\bf q}\cdot {\bf n}  =  q_1n_1   + \ldots+ q_d n_d $. We shortly write
\begin{subequations}
\begin{align}
\label{sum-d}
\sum_{{\bf n}} & =  \sum_{n_1=-\infty}^\infty \cdots \sum_{n_d=-\infty}^\infty \\
\label{dq}
\int d {\bf q} &=  \int_0^{2\pi}  \frac{dq_1}{2\pi}\, \cdots
\int_0^{2\pi}\frac{dq_d}{2\pi}       
\end{align}
\end{subequations}
We also define 
\begin{subequations}
\begin{align}
\label{Psi-def}
\Psi_d(s, {\bf q})   & = \left[s - 2 \ii \sum_{a=1}^d\cos
  q_a\right]^{-1}\\
\label{Phi-def}
\Phi_d(s)              &= \left[\int d {\bf q}\,\Psi_d(s, {\bf q})
  \right]^{-1}
\end{align}
\end{subequations}
Equation \eqref{Phi-def} expresses $\Phi_d(s)$ through a $d-$fold integral. This multiple integral can be 
reduced to a single integral involving a Bessel function:
\begin{equation}
\label{Phi-Bessel}
\Phi_d(s) = \left\{\int_0^\infty
du\,e^{-us}\left[J_0(2u)\right]^d\right\}^{-1}
\end{equation}
This is established using \eqref{Psi-def}--\eqref{Phi-def} and the well-known integral representation 
\begin{equation}
\label{Bessel}
J_0(z) = \int_0^{2\pi} \frac{dq}{2\pi}\, e^{2z \ii \cos q}
\end{equation}
 of the Bessel function.

Applying  the Laplace-Fourier transform to the governing equations we arrive at
\begin{equation}
\label{SS0:d}
S(s, {\bf q}) = \left[1 +  \Gamma \widehat{S}_{\bf 0}(s)\right]
\Psi_d(s, {\bf q}) 
\end{equation}
From  the definition \eqref{Ssq-d}, we obtain 
\begin{equation}
\label{S0:d-int}
\widehat{S}_{\bf 0}(s) = \int d {\bf q}\,\,S(s, {\bf q})
\end{equation}
Using \eqref{SS0:d} and \eqref{S0:d-int} we find that 
\begin{equation}
\label{S0:d}
\widehat{S}_{\bf 0}(s) = \left[\Phi_d(s) - \Gamma\right]^{-1} 
\end{equation}
with $\Phi_d(s)$ defined in \eqref{Phi-def}. Therefore \eqref{SS0:d} becomes 
\begin{equation}
\label{S-d-exact}
S(s, {\bf q}) = \frac{\Phi_d(s)}{\Phi_d(s) - \Gamma}\, \Psi_d(s, {\bf q}) 
\end{equation}
Similarly to one dimension, the growth of the total  average number of bosons is governed by the differential equation
\begin{equation}
\label{N-d:S0}
\frac{d^2 N}{dt^2} = (2\Gamma)^2\, |S_{\bf 0}(t)|^2
\end{equation}

\subsection{Two Dimensions}
\label{subsec:2d}

For the square lattice, the integral in Eq.~\eqref{Phi-Bessel} admits a representation through a simpler integral (see \cite{Hughes}) 
\begin{equation}
\label{Phi-2}
\Phi_2(s) = \frac{\pi}{2}\left[\int_0^{\pi/2}\frac{d\theta}{\sqrt{s^2+16\sin^2\theta}}\right]^{-1}
\end{equation} 
The integral in \eqref{Phi-2} can be expressed through complete elliptic integrals, but the form \eqref{Phi-2} suffices for us. As in one dimension where $\Phi_1(s) = \sqrt{s^2+4}$, the function $\Phi_2(s)$ is a monotonically increasing function of $s$. The key difference from the one-dimensional case is that $\Phi_2(0)=0$ and hence $\widehat{S}_{\bf 0}(s)$ has a pole for any $\Gamma>0$. The pole is simple and therefore
\begin{equation}
\label{S0:2-asymp}
S_{\bf 0}(t)\simeq \alpha_2(\Gamma)\,e^{C_2(\Gamma) t}
\end{equation}
when $t\gg 1$, with $C_2(\Gamma)$ and $\alpha_2(\Gamma)$ implicitly determined from
\begin{equation}
\label{CA2:def}
\Phi_2[C_2(\Gamma)] = \Gamma, \quad \frac{1}{\alpha_2(\Gamma)} = \frac{d \Phi_2}{ds}\Big|_{s=C_2(\Gamma)}
\end{equation}
Combining \eqref{N-d:S0} and \eqref{S0:2-asymp} we obtain
\begin{equation}
\label{N:2-asymp}
N(t)\simeq A_2(\Gamma)\,e^{2C_2(\Gamma) t}, \quad 
A_2(\Gamma) = \left[\frac{\Gamma \alpha_2(\Gamma)}{C_2(\Gamma)}\right]^2
\end{equation}

Using the asymptotic
\begin{equation}
\label{Phi-2-small}
\Phi_2(s) = \frac{2\pi}{\ln(16/s)} +O\left[\frac{s^2}{\ln(1/s)}\right]
\end{equation} 
and \eqref{CA2:def} we find $C_2(\Gamma)\simeq 16\,e^{-2\pi/\Gamma}$ when $0<\Gamma\ll 1$. Thus when $\Gamma\ll 1$, the growth remains exponential but has a very small amplitude in the exponent:
\begin{equation}
\lim_{t\to \infty} \frac{\ln N(t)}{t} = 32\,e^{-2\pi/\Gamma}
\end{equation}


\begin{figure}
\centering 
\includegraphics[width=0.66\textwidth]{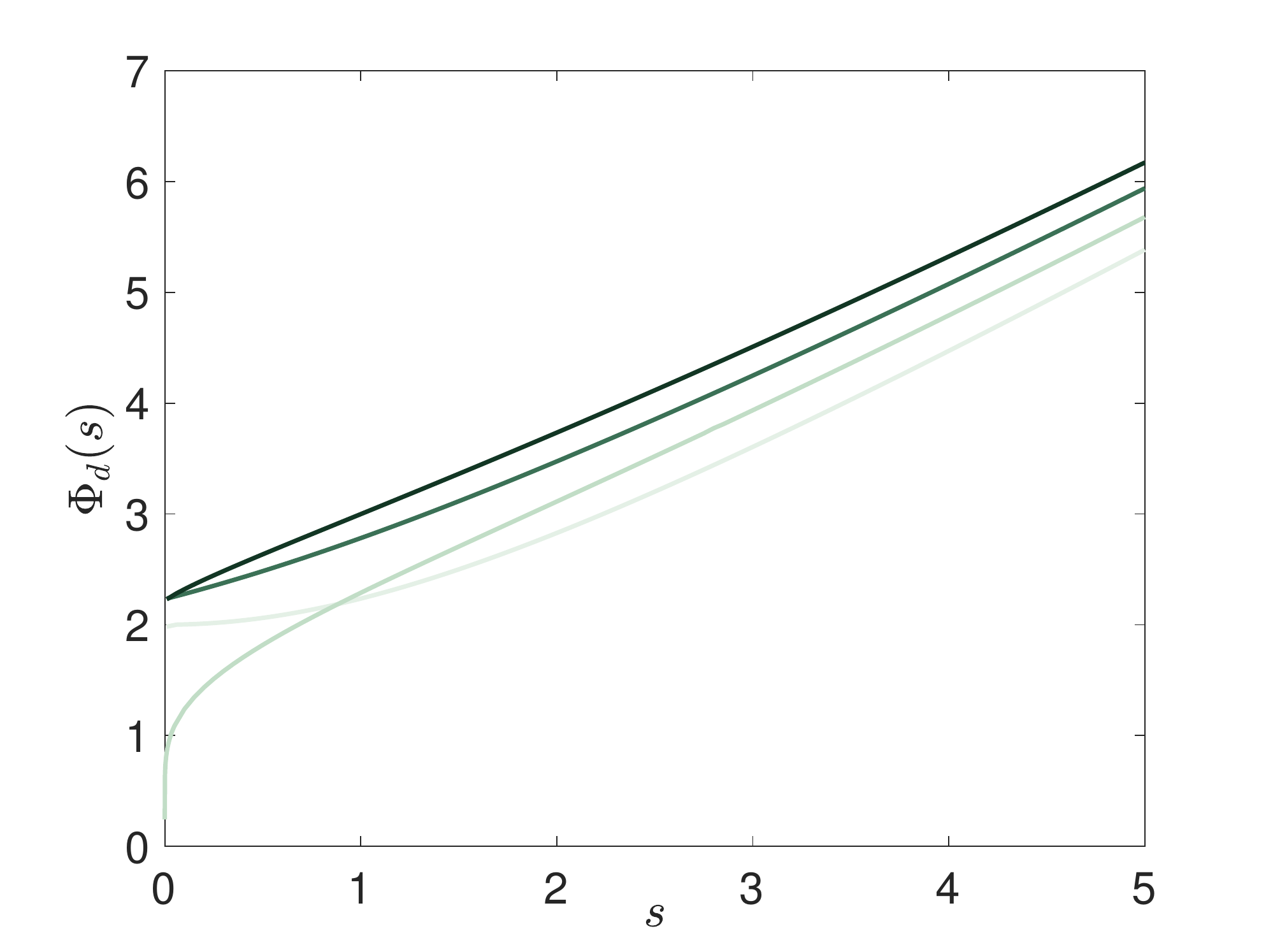}
\caption{The functions $\Phi_d(s)$ for $d=1,2,3,4$. The curves are asymptotically parallel and go from light to dark (or bottom to top when $s\gg 1$) as $d$ increases, see \eqref{Phi-d-large}. }
\label{Fig:F1234}
\end{figure}

\subsection{High Dimensions}

The functions $\Phi_d(s)$ are monotonically increasing functions of $s$ in all dimensions and $\Gamma_d=\Phi_d(0)>0$ in all dimensions apart from $d=2$, see Fig.~\ref{Fig:F1234}. We already know $\Gamma_1=2$ and  $\Gamma_2 =0$. The critical input rate in three dimensions can be expressed via the Euler's gamma function
\begin{equation}
\label{gamma-3}
\Gamma_3 = \frac{3}{\pi}\left[\Gamma\!\left(\frac{5}{6}\right)\cdot \Gamma\!\left(\frac{2}{3}\right)\right]^2
\end{equation}
The critical input rate in four dimensions
can be expressed through a Meijer G-function (see \cite{Bateman,Meijer} for the definition of the Meijer G-functions):
\begin{equation}
\Gamma_4  = \frac{4\pi}{\text{MeijerG}\!\left[\{\{1,1\}, \{1, 1\}\}, \{\{\frac{1}{2}, \frac{1}{2}\}, \{\frac{1}{2}, \frac{1}{2}\}\}, 1\right]}
\end{equation}
Here are a few numerical values of the critical input rate $\Gamma_d=\Phi_d(0)$:
\begin{equation}
\begin{split}
\Gamma_1 & = 2 \\
\Gamma_2 & = 0 \\
\Gamma_3 & = 2.23104529048681\ldots \\
\Gamma_4 & = 2.2155052677337\ldots  \\
\Gamma_5 & = 2.65391372999969\ldots  \\
\Gamma_6 & = 2.831160419094839\ldots  \\
\Gamma_7 & = 3.07421878003498\ldots
\end{split}
\end{equation}
Note oscillations depending on the parity of $d$. For small $d$ these oscillations are particularly notable and only for $d\geq 4$ the critical input rates $\Gamma_d$ exhibit a monotonic growth when the spatial dimension increases. Using  \eqref{Phi-Bessel}  and the small $x$ expansion of the Bessel function
\begin{equation}
\label{Bessel:small}
J_0(x)=1-\frac{x^2}{4}+\ldots
\end{equation}
one can determine the large $d$ behavior of the critical input rate:
\begin{equation}
\Gamma_d \simeq \sqrt{\frac{4d}{\pi}} 
\end{equation}

Combining \eqref{Phi-Bessel}  and \eqref{Bessel:small} one also finds the large $s$ asymptotic of $\Phi_d(s)$: 
\begin{equation}
\label{Phi-d-large}
\Phi_d(s) = s + \frac{2d}{s}+\ldots
\end{equation}
This  large $s$ behavior helps to identify the curves in Fig.~\ref{Fig:F1234}. It also shows that the effective multiplication rate $C_d(\Gamma)$ appearing in \eqref{Nd:all}, see also \eqref{N:d-asymp} below, behaves as 
\begin{equation}
\label{Cd-large}
C_d(\Gamma) = \Gamma - \frac{2d}{\Gamma}+\ldots
\end{equation}
when $\Gamma\gg 1$. 

Below we show that $N(t)$ is indeed described by \eqref{Nd:all} in the large time limit. 

\subsubsection{The supercritical regime: $\Gamma>\Gamma_d$.}

In this region in $d\geq 3$ dimensions, the Laplace transform $\widehat{S}_{\bf 0}(s)$ has a simple pole at $s=C_d(\Gamma)>0$ implying an exponential long time behavior
\begin{equation}
\label{S0:d-asymp}
S_{\bf 0}(t)\simeq \alpha_d(\Gamma)\,e^{C_d(\Gamma) t}
\end{equation}
with $C_d(\Gamma)$ and $\alpha_d(\Gamma)$ implicitly determined from
\begin{equation}
\Phi_d[C_d(\Gamma)] = \Gamma, \quad \frac{1}{\alpha_d(\Gamma)} = \frac{d \Phi_d}{ds}\Big|_{s=C_d(\Gamma)}
\end{equation}
Thus the average total number of bosons exhibits an exponential growth 
\begin{equation}
\label{N:d-asymp}
N(t)\simeq A_d(\Gamma)\,e^{2C_d(\Gamma) t}, \quad 
A_d(\Gamma) = \left[\frac{\Gamma \alpha_d(\Gamma)}{C_d(\Gamma)}\right]^2
\end{equation}

\subsubsection{The critical case: $\Gamma=\Gamma_d$.}
At the critical point, the Laplace transform  $\widehat{S}_{\bf 0}(s)$ has a simple pole at the origin, namely $\widehat{S}_{\bf 0}(s)\simeq 1/(\beta_d s)$, from which
\begin{equation}
\label{S0:d-crit}
S_{\bf 0}(t)\simeq \frac{1}{\beta_d}\,, \quad \beta_d =  \frac{d \Phi_d}{ds}\Big|_{s=0}
\end{equation}
leading to
\begin{equation}
\label{Nd-crit}
N(t)\simeq  B_d^\text{crit}\, t^2, \quad B_d^\text{crit}=2(\Gamma_d/\beta_d)^2
\end{equation}
In three dimensions
\begin{equation*}
\beta_3=(\Gamma_3)^2\int_0^\infty
du\,u\left[J_0(2u)\right]^3
\end{equation*}
Using identities
\begin{equation*}
\int_0^\infty du\,u\left[J_0(2u)\right]^3 = \frac{1}{3}\int_0^\infty du\,\left[J_1(u)\right]^3 = \frac{1}{2\pi \sqrt{3}}
\end{equation*}
and \eqref{gamma-3} we express $\beta_3$ through the Euler's gamma function
\begin{equation}
\label{beta-3}
\beta_3 = \frac{3\sqrt{3}}{2\pi^3}\left[\Gamma\left(\frac{5}{6}\right)\cdot \Gamma\left(\frac{2}{3}\right)\right]^4=0.45737905988493\ldots
\end{equation}
Using \eqref{gamma-3}  and \eqref{beta-3} we see that in three dimensions 
\begin{equation}
B_3^\text{crit}= \frac{8}{27}\left[\frac{\pi}{\Gamma\!\left(\frac{5}{6}\right)\cdot \Gamma\!\left(\frac{2}{3}\right)}\right]^4
= 5.28751612048\ldots
\end{equation}

Here are numerical values for $\beta_d$ in the range $4\leq d\leq 13$:
\begin{equation}
\begin{split}
\beta_4 & = 1.89180068997\ldots  \\
\beta_5 & = 0.58095222806\ldots  \\
\beta_6 & = 0.67480419748\ldots  \\
\beta_7 & = 0.61631056140\ldots\\
\beta_8 & = 0.63335480913\ldots\\
\beta_9 & = 0.62600792587\ldots\\
\beta_{10} & = 0.62977975271\ldots\\
\beta_{11} & = 0.62908952670\ldots\\
\beta_{12} & = 0.63021496135\ldots\\
\beta_{13} & = 0.63047523664\ldots
\end{split}
\end{equation}
Oscillations depending on the parity of $d$ are even more pronounced than for  the critical input rate $\Gamma_d$ and only after $d\geq 11$ the quantities $\beta_d$ become increasing functions of the dimensionality. Using \eqref{Phi-Bessel}  and \eqref{Bessel:small} one establishes the saturation 
\begin{equation*}
\lim_{d\to\infty} \beta_d = \frac{2}{\pi}
\end{equation*}
of the quantities $\beta_d$ in the large $d$ limit. 

\subsubsection{The subcritical domain: $\Gamma<\Gamma_d$.}

Below the critical point,  the Laplace transform $\widehat{S}_{\bf 0}(s)$ has two simple poles at $s=\pm \ii \sigma_d(\Gamma)$ 
\begin{equation}
\label{Phi-Bessel-Gamma}
\int_0^\infty du\,\cos\{\sigma_d(\Gamma)\,u\}\,\left[J_0(2u)\right]^d = \frac{1}{\Gamma}
\end{equation}
The contribution of these two poles into the inverse Laplace transform leads to the oscillating leading behavior similar to \eqref{S0:1-osc}, namely
\begin{equation}
\label{S0:d-osc}
S_{\bf 0}(t)\simeq \Gamma^{-1} \sqrt{B_d(\Gamma)}\,\sin\!\left[t \sigma_d(\Gamma)\right]
\end{equation}
Using \eqref{N1:S0} and \eqref{S0:d-osc} we arrive at $N(t)\simeq B_d(\Gamma)\, t^2$ as stated in \eqref{Nd:all}.

\section{Density profile in high dimensions}
\label{sec:d-density}

Using
\begin{equation*}
\rho_{\bf 0}(t) = 2\Gamma \int_0^t dt'\, |S_0(t')|^2
\end{equation*}
and already established results for $S_{\bf 0}(t)$ we arrive at
\begin{equation}
\label{N0-d:all}
\rho_{\bf 0}(t)\simeq
\begin{cases}
\Gamma^{-1} A_d(\Gamma) C_d(\Gamma)\,e^{2C_d(\Gamma)\, t}  & \Gamma>\Gamma_d \\
\Gamma_d^{-1} B_d^\text{crit}\, t           & \Gamma = \Gamma_d  \\
\Gamma^{-1}B_d(\Gamma)\, t               & \Gamma < \Gamma_d  \\
\end{cases}
\end{equation}
describing the asymptotic growth of the density at the origin when $d\geq 3$. Comparing with \eqref{Nd:all} we see that in the supercritical regime the fraction of bosons at the origin approaches  $\Gamma^{-1} C_d(\Gamma)$ as $t\to\infty$. Thus we have an intriguing spatial counterpart of Bose-Einstein condensation.

Let us analyze the entire density profile. We limit ourselves to the supercritical regime. We begin with the square lattice. In this situation $\Gamma_2=0$, so there is only supercritical regime.  We must solve
the following equation
\begin{equation}
\label{S-mn}
 \frac{d  S_{m,n}}{dt} = \ii [S_{m+1,n} + S_{m-1,n} + S_{m,n+1} +  S_{m,n-1}] +  \Gamma \delta_{m,0}\delta_{n,0} S_{m,n}
\end{equation}

We know that $S_{0,0}(t)\sim e^{C_2(\Gamma)\, t}$, see section \ref{subsec:2d}. This together with the behavior of the density in the supercritical regime in one dimension suggest to seek the general solution of \eqref{S-mn} in the form
\begin{equation}
S_{m,n}(t)=F_{m,n}\,e^{C_2(\Gamma)\, t}
\end{equation}
Plugging this ansatz into \eqref{S-mn} we obtain
\begin{equation}
\label{F-mn}
F_{m+1,n} + F_{m-1,n} + F_{m,n+1} +  F_{m,n-1} =  - \ii  C_2(\Gamma) F_{m,n} + \ii \Gamma \delta_{m,0}\delta_{n,0} F_{m,n}
\end{equation}
Taking into  account that $F_{1,0}=F_{-1,0}=F_{0,1}=F_{0,-1}$ due to symmetry we simplify \eqref{F-mn} to 
\begin{equation}
\label{F-00}
4F_{1,0} =   \ii  [\Gamma- C_2(\Gamma)]F_{0,0}
\end{equation}
at the origin. Thus the density in the nearest-neighbor sites of the origin is 
\begin{equation}
\label{density-nn}
\frac{\rho_{1,0}}{\rho_{0,0}} = \left[\frac{\Gamma- C_2(\Gamma)}{4}\right]^2
\end{equation}
Since $\Gamma- C_2(\Gamma)\simeq 4/\Gamma$ for $\Gamma\gg 1$, see \eqref{Cd-large}, we obtain 
\begin{equation}
\label{density-nn-asymp}
\frac{\rho_{1,0}}{\rho_{0,0}} \simeq \Gamma^{-2}\qquad\text{when}\quad \Gamma\to\infty
\end{equation}

\begin{figure}[t]
\centering
\includegraphics[width=0.77 \textwidth]{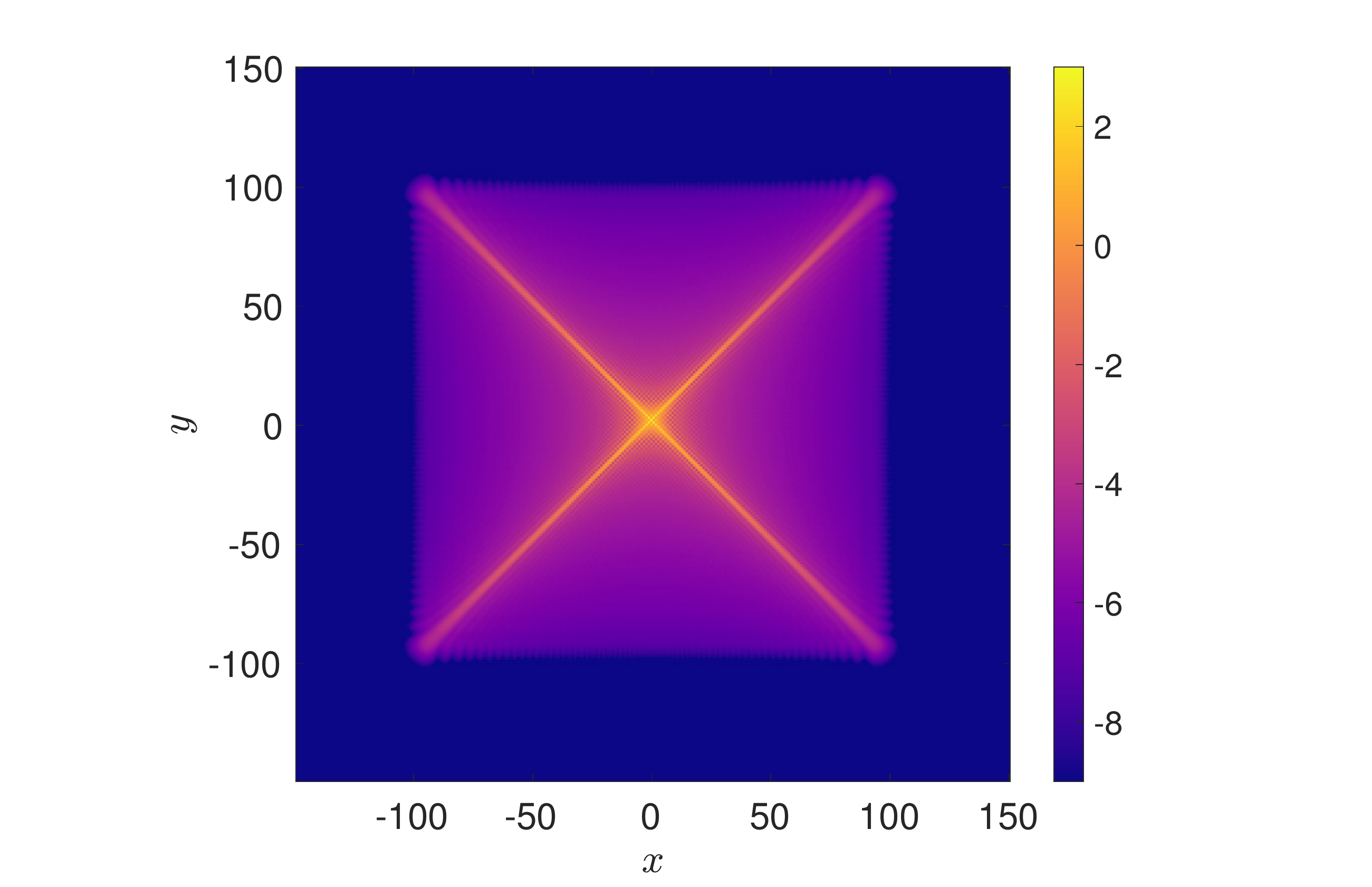}
\caption{The density on a $300 \times 300$ square lattice at time $t=50$ for $\Gamma=1$. The color axis is logarithmic to resolve any structure away from the exponentially large density on the (co)-diagonal.}
\label{fig:2Ddens}
\end{figure}

Simulations indicate that $F_{m,n}$ is remarkably anisotropic, namely it quickly vanishes outside the diagonal $m=n$ and co-diagonal $m=-n$; along the diagonal and co-diagonal, $F_{m,n}$ is also quickly decaying when the separation from the origin increases; see Fig.~\ref{fig:2Ddens}. Let us try to understand these properties using approximate techniques. Consider for concreteness the positive quadrant $\{(m,n)|\,m>0, \, n>0\}$ and assume that $C_2\ll 1$. This latter (technical) assumption insures that the decay along the diagonal is slow (albeit still exponential as we shall derive) and this allows to employ continuum methods. Thus we make  the ansatz
\begin{equation}
\label{FiF}
F_{m,n} = \ii^{m+n}\,\Phi(m,n)
\end{equation}
and treat $\Phi(m,n)$ as a function of continuous variables --- this is justified when $m\gg 1$ and  $n\gg 1$ if additionally $C_2\ll 1$. Plugging the ansatz \eqref{FiF} into \eqref{F-mn} we obtain
\begin{equation*}
\Phi(m-1,n)-\Phi(m+1,n)+ \Phi(m,n-1)-\Phi(m,n+1) = C_2 \Phi(m,n)
\end{equation*}
which becomes 
\begin{equation}
\label{mn-Phi}
(\partial_m+\partial_n)\Phi+\tfrac{1}{2}C_2\Phi=0
\end{equation}
in the continuum approximation. Introducing the variables
\begin{equation}
u=m-n, \quad v = m+n
\end{equation}
we recast \eqref{mn-Phi} into $\partial_v \Psi+\tfrac{1}{4}C_2\Psi=0$ for $\Psi(u,v)=\Phi(m,n)$. Thus
\begin{equation}
\label{uv-Psi}
\Psi(u,v)=\chi(u)\,e^{-C_2 v/4}
\end{equation}
with arbitrary $\chi(u)$. We haven't computed $\chi(u)$, but assuming  it quickly vanishes as $|u|$ increases, this would explain numerical observations (shown in Fig.~\ref{fig:2Ddens}) that the density is mostly concentrated along the diagonal and co-diagonal. Moreover, \eqref{uv-Psi} supports the exponential decay along the diagonal and co-diagonal which we observe numerically. 

An exact analysis of \eqref{F-mn} is possible. Applying the Fourier transform 
\begin{equation}
\label{F-Fourier}
\mathcal{F}(p,q) =  \sum_{m=-\infty}^\infty \sum_{n=-\infty}^\infty F_{m,n}\,e^{- \ii pm}\,e^{- \ii qn}
\end{equation}
to \eqref{F-mn} we find a neat expression
\begin{equation}
\label{Fpq}
\mathcal{F}(p,q) = \frac{\Gamma F_{0,0}}{C_2(\Gamma) - 2\ii (\cos p + \cos q)}
\end{equation}
leading to the integral representation 
\begin{equation}
\label{Fmn-sol}
F_{m,n} = \frac{1}{(2\pi)^2}\int_0^{2\pi} dp \int_0^{2\pi} dq\, 
\frac{\Gamma F_{0,0}\, e^{\ii pm}\,e^{\ii qn}}{C_2(\Gamma) - 2\ii (\cos p + \cos q)}
\end{equation}
Euclidean versions of the integrals similar to those appearing in \eqref{Fpq}--\eqref{Fmn-sol} are known as the lattice Green functions, see e.g. \cite{Hughes,Joyce03,Guttmann,Ray} and references therein. These lattice Green functions can be expressed through special functions for a very few lattices, and even when this is possible the resulting formulas tend to be extremely complicated. For instance, for the square lattice the corresponding Green function has been expressed \cite{Ray} through the hypergeometric function $_{5}F_4$, but extracting asymptotic behaviors from these exact formulas appears rather difficult.

Generally on the $\mathbb{Z}^d$ lattice we have $S_{\bf n}(t)=F_{\bf n}\,e^{C_d(\Gamma)\,t}$ in the supercritical regime with
\begin{equation}
\label{Fnd-sol}
F_{\bf n} = \int\! d{\bf q}\,\,
\frac{\Gamma F_{\bf 0}\, e^{\ii {\bf q}\cdot {\bf n}}}{C_d(\Gamma) - 2\ii \sum_{1\leq a\leq d} \cos q_a}
\end{equation}
where $\int\! d{\bf q}$ is defined via \eqref{dq}. The density $\rho_{\bf 0}'$ in the nearest-neighbor sites of the origin is given again by a rather simple formula 
\begin{equation}
\label{density-nn-d}
\frac{\rho_{\bf 0}'}{\rho_{\bf 0}} = \left[\frac{\Gamma- C_d(\Gamma)}{2d}\right]^2
\end{equation}

\section{Microscopic model}
\label{sec:model}

We have given exact analytic expressions, confirmed by numerical simulations, that show there is exponential growth of the number of particles above a certain critical injection rate $\Gamma_c$. But, how can we  rationalize this? The Lindblad equation~\eqref{LindbladB} describes the evolution of the ensemble average of a set of quantum trajectories, each evolving under a Hamiltonian $H$, while being subjected to a Poisson process that injects particles with an average rate $\Gamma$. If injection events occur at a fixed rate, it appears impossible to get an exponential growth. Let's try to resolve this apparent paradox by constructing a microscopic model that gives rise to Lindbladian dynamics~\eqref{LindbladB}. For Lindbladian~\eqref{Lindblad:origin}, the dissipative part of the Lindblad equation~\eqref{LindbladB} is quadratic. As a consequence, we expect that there exists a simple bilinear Hamiltonian for the system and reservoir that gives rise to the dissipative Markovian dynamics once the reservoir is traced out. Moreover, everything can be understood on the level of the Heisenberg equations of motion for the bosonic field operators. Before we investigate those equations for the closed system+reservoir dynamics, let's have a look at the equations of motion for the Lindblad dynamics. Recall that the Lindblad equation for observables in the Heisenberg picture is given by
\begin{equation}
\partial_t O= \ii [H, O] + 2L^\dagger O L-\{L^{\dagger}L, O\}.
\end{equation}
Consequently for $O=b_i$, we find
\begin{equation}
\label{eq:Heis_b}
\partial_t b_i=\ii [H,b_i]+\Gamma \delta_{i,0}b_i,
\end{equation}
which is simply equation~\eqref{S-dual}. The most naive model would correspond to 
\begin{equation}
H_{\rm naive}= H+ \sum_i \omega_i a^\dagger_i a_i +\sum_i g_i (b^\dagger_0 a_i+a_i^\dagger b_0),
\end{equation}
where $a_i$ are bosonic bath annihilation operators. By carefully choosing the spectrum $\omega_i$ and the coupling $g_i$, one can make sure the reduced dynamics becomes Markovian, i.e. one needs to guaranty that the noise spectral density is white such that expression~\eqref{eq:spectrumnoise} holds. The interaction term annihilates particles from the bath and puts them in the system. While it seems natural that this would lead to the Lindbladian~\eqref{Lindblad:origin}, when we start with a bath that contains a lot of particles, we will now show it does not. The Heisenberg equations of motion for the bath degrees of freedom are
\begin{equation}
\ii \partial_t a_i = \omega_i a_i + g_i b_0.
\end{equation}
This can readily be solved and we can simply substitute its solution into the Heisenberg equation of motion of the system operators, yielding
\begin{equation}
 \partial_t b_i=\ii[H,b_i ]- \delta_{i,0}\int^t_0{\rm d}s \sum_j g_j^2 e^{-\ii \omega_j (t-s)} b_i(s)-\ii \delta_{i,0} \sum_j g_j e^{-\ii \omega_j t} a_j(0)
\end{equation}
Taking the thermodynamic limit for the bath degrees of freedom, such that 
\begin{equation}
\sum_j g_j^2 e^{-\ii \omega_j t} \rightarrow \Gamma \delta(t-s),
\label{eq:spectrumnoise}
\end{equation}
we arrive at
\begin{equation}
 \partial_t b_i=\ii[H,b_i ]- \Gamma \delta_{i,0} b_i+\delta_{i,0} f(t)
\end{equation}
Here $f(t)$ is a stochastic driving term that depends on the initial state of the reservoir and the exact nature of the couplings $g_i$. If we start with a reservoir that has $N$ particles, the stochastic driving term $f\sim \sqrt{N}$. It will pump particles into the system, however, the self-interaction that the system picks up by interacting with the bath has the opposite sign of that in the Lindblad equation~\eqref{eq:Heis_b}, i.e. it removes particles from the system rather than injecting them. Only the stochastic drive will inject particles. As a consequence, $H_{\rm naive}$ does not describe the Lindblad dynamics for the system. To generate the desired dynamics, we need to flip the sign of the self-interaction term. This can be done by swapping the bath creation and annihilation operators:
\begin{equation}
\label{eq:bog}
H_{\rm Matthew}= H+ \sum_i \omega_i a^\dagger_i a_i +\sum_i g_i (b^\dagger_0 a_i^\dagger+a_i b_0)
\end{equation}
Going through the same exercise as before, we derive the equation of motion 
\begin{equation}
 \partial_t b_i=\ii[H,b_i ]+ \Gamma \delta_{i,0} b_i+\delta_{i,0} f(t)
 \label{eq:heispair}
\end{equation}
We can always initialize the bath in the vacuum state with no particles. The latter would make $\left<f\right>=0$ leading to Eq.~\eqref{eq:Heis_b}. Thus, we can think of the Lindbladian~\eqref{Lindblad:origin} in terms of $H_\text{Matthew}$ which instead of removing a particle from the reservoir and putting it into the system creates a pair of particles, one in the reservoir and one in the system. The term $\Gamma b_i \delta_{i,0}$ in Eq.~\eqref{eq:heispair} triggered by the input now comes with positive sign leading to the rich get richer, or quantum Matthew effect. The total particle number is not conserved, which makes it possible to sustain this exponential growth. Moreover, the transition from quadratic to exponential growth can be interpreted as a transition from a dynamically stable to a dynamically unstable system~\eqref{eq:bog}. By itself, this Hamiltonian is unphysical in the dynamically unstable regime as it is also thermodynamically unstable, i.e. the ground state has unbounded energy. However, the Hamiltonian can always be thought of as an effective Hamiltonian arrising as the Bogoliubov description of a more complex problem. Examples of the latter, in which similar physics appears, include the dynamical Casimir effect~\cite{Marios} and parametric instabilities in driven systems~\cite{Marin}. We also notice that an open {\em classical} system of non-interacting random walks multiplying on a `fertility' site is somewhat similar to the open quantum system described by the Hamiltonian \eqref{eq:bog} and it exhibits similar behaviors (see \cite{Saclay19} and references therein).

\section{Conclusions}
\label{sec:concl}

Open quantum systems driven by a localized source demonstrate the  dramatic difference 
between fermions and bosons.  
In the case of identical fermions, the behavior is rather predictable \cite{KMS19,Kollath}, 
e.g., the total number of fermions grows linearly in time and only the dependence 
of the fermion growth rate on  the input rate is unusual, viz. it decays as $1/\Gamma$ 
when the source becomes strong which is a manifestation  of the quantum Zeno effect. 
For bosons, the growth is super-linear: quadratic in time when $\Gamma<\Gamma_d$ 
and exponential when $\Gamma > \Gamma_d$. The tendency of identical bosons to bunch is well-known in the context of closed quantum systems, with Bose-Einstein condensation being the beautiful illustration of this phenomenon. Another way to express the exponential growth in our open quantum system of identical bosons is to think about it in terms of the microscopic `Matthew' Hamiltonian [see~\eqref{eq:bog}] leading to rich get richer, or the quantum Matthew effect. Our work indicates that in the context of open quantum systems the difference between fermions and bosons can be more spectacular than for closed quantum systems. 

We analyzed the asymptotic behavior of the average total number of injected bosons. Needless to say, one would like to probe the behavior of higher cumulants, and to study  the entire distribution function $P(N,t)$ or the full counting statistics. A self-averaging behavior is anticipated in the case of identical fermions, so the average total number of injected fermions provides an asymptotically complete description. In the case of non-interacting bosons, self-averaging or the lack of it is an interesting issue. At the transition point $\Gamma=\Gamma_d$, the fluctuations could be especially pronounced. 

Another obvious extension is related to the spatial distribution of bosons. We determined the density profile in one dimension. In higher dimensions, particularly in the supercritical regime, we derived exact integral representations. Extracting asymptotic behaviors from these integral representation is an intriguing challenge since the numerically observed spatial distribution is strikingly anisotropic (see Fig.~\ref{fig:2Ddens}).  We also emphasize that in the supercritical regime the fraction of bosons at the origin approaches to a constant, namely to $\Gamma^{-1}C_d(\Gamma)$. This could be interpreted as Bose-Einstein condensation in a spatially localized state.

An open quantum system of bosons driven by a localized source has been previously studied in Ref.~\cite{Spohn10} where a possibility of an exponential growth has been discovered. Our analysis is much more detailed as the lattice system appears to be more amenable to analytical work. Intriguingly, the linear in time growth occurs only when $t\ll 1$, while for $t\gg 1$ even in the sub-critical regime the growth is quadratic. It would be interesting to find an intuitive understanding of the quadratic $N\propto t^2$ growth law.

An intriguing feature of our open system of bosons is a phase transition occurring for all $d\ne 2$. The two-dimensional case is special in numerous problems where the mathematical framework relies on the Laplacian, e.g. in {\em classical} diffusive systems driven by a localized source the two-dimensional behaviors could be subtle, see e.g. \cite{PLK:SEP,Darko:SEP}. Still, the peculiar role of $d=2$ in the present situation is striking. We established this special behavior for the square lattice, so it would be interesting to probe its validity for other two-dimensional lattices. 

\section*{Acknowledgments}
PLK thanks the Institut de Physique Th\'eorique and the Los Alamos National Laboratory for hospitality.  KM is grateful to Michel Bauer, Jean-Marc Luck and  Toma\v{z} Prosen for illuminating discussions. DS acknowledges support from the FWO as post-doctoral fellow of the Research Foundation -- Flanders.

\end{document}